\documentclass[sigconf,authordraft=False,anonymous=False,nonacm=True]{acmart}
\AtBeginDocument{%
  }

\setcopyright{rightsretained}
\copyrightyear{2024}
\acmYear{2024}
\acmDOI{arXiv:2411.05985}
\acmConference[arXiv]{arXiv}{Dec 13,
  2024}{Washington, DC}
\acmISBN{arXiv:2411.05985}

\usepackage{subcaption}

\begin{document}

\title{Generating Fearful Images: Investigating Potential Emotional Biases in Image-Generation Models}

\author{Maneet Mehta}
\affiliation{%
  \institution{Yale University}
  \city{New Haven, CT}
  \country{USA}
}

\author{Cody Buntain}
\affiliation{%
  \institution{University of Maryland}
  \city{College Park, MD}
  \country{USA}
}

\renewcommand{\shortauthors}{Mehta and Buntain}

\begin{abstract}
This paper examines potential biases and inconsistencies in the emotions evoked by images produced by generative artificial intelligence (AI) models and their potential bias toward negative emotions. 
We assess this bias by comparing the emotions evoked by an AI-produced image to the emotions evoked by prompts used to create those images.
After developing and validating automated methods for emotion recognition across modalities, we examine correlations in the prevalence of emotions across text and images and measure the degree to which generative AI models tend to over-represent specific emotions in the resulting images.
Findings indicate that AI-generated images from Stable Diffusion models are biased towards producing images that evoke fear, regardless of the original prompt, as metrics show a significant over-representation of that emotion compared to five other emotions. 
We extend this analysis to a more recent enterprise-level models, such as ChatGPT and Gemini, and find similar results, suggesting a systemic bias rather than one present only in a single model.
While certain limitations in the alignment of emotions across modalities limit this work, the emotional skew we find in generative models is consistent with an over-representation of fearful content in training data, and this bias could amplify negative affective content in digital spaces further, perpetuating its prevalence and impact.  
\end{abstract}



\received{30 November 2024}

\maketitle

\section{Introduction}

The information environment increasingly relies on visual media, with visual-first platforms like YouTube, Instagram, Pinterest, and TikTok comprising four of the top five most-used social media platforms in the US \citep{pew:2024}.
In studying these online spaces, the academic consensus has increasingly solidified around two findings: 1) the presence of visual media increases online engagement \citep{10.1177/1065912918786805.2019,10.1177/1940161220968534.2021}, and 2) highly emotional content---especially negative---also increases this engagement \citep{10.1073/pnas.2024292118.2021,10.1126/sciadv.ade9231}.
At the same time, generative AI models' ever-growing hunger for training data has lead to increased  use of social media and creator-content as training data for modern large- and visual-language models.
Together, these factors have the potential to produce an ``unvirtuous cycle'' between generative AI models and the information environment from which large volumes of training data are collected.
In this cycle, content creators produce more emotionally evocative visual media to gain more engagement, generative AI systems then over-represent negative emotions in the media they produce, which then feeds back into the information environment, receives more engagement, gains a larger share of visual media we see, and gets ingested in the next round of training generative-AI systems.

This paper begins an investigation of this unvirtuous cycle by developing scalable emotion-recognition methods for images and comparing these emotions to those present in the underlying prompts. We demonstrate that fine-tuned computer vision models—particularly Google’s Vision Transformer (ViT)—effectively identify emotions in images. Using a fine-tuned ViT model coupled with a state-of-the-art model for emotion recognition in text \citep{10.1109/icassp49357.2023.10095597,10096864}, we compare the emotions present in a text-based generative-AI prompt to the emotions evoked by the produced image.
Ideally, the distributions of emotion present in generated images should mirror the distribution in the underlying prompt used to generate this image.
We suspect, however, that this unvirtuous cycle between generative AI and the information environment instead produces an anti-social outcome: that \textbf{generative-AI systems are biased toward producing images that evoke negative emotions regardless of the underlying prompt.} We measure this using true- and false-positive rates per emotion, allowing us to distinguish systematic over-generation of an emotion 

Results show that, for at least one popular generative AI text-to-image model (i.e., Stable Diffusion), the emotions evoked by an image do appear more negative than their source prompts, where the Stable Diffusion images from the dataset substantially over-represent fear, whereas the prompts primarily show excitement. Image models appear better at both replicating fear and also generating it even when not directly prompted to, indicating a structural propensity towards fear.
To extend this analysis beyond one generative model, we also sample prompts from the DiffusionDB dataset and use OpenAI's GPT-Image-1.5 model to generate new images using enterprise-level models; our results are consistent for this subset as well, where GPT images also over-represent negative emotions (fear) relative to the source prompts.
These results yield evidence of the concerning hypothesis above, namely that a AI-based image-generation models may nudge their users and audiences toward negative emotions.
This finding is particularly concerning given the state of our modern information environment, where we know negative emotions are contagious \citep{doi:10.1073/pnas.1320040111}.
Hence, we advocate for a multidisciplinary approach to better align AI emotion recognition with psychological insights and address potential biases in generative AI outputs across digital media.


%
%
\section{Related Work}

%

This investigation builds on work from two main communities: scholars of fairness and bias in AI systems, and computational social science scholars of the information environment.
From the first community---which includes many aspects of AI evaluation and auditing---major concerns have arisen over the potential for unintended harms caused by AI systems, such as the unvirtuous cycle we outline.
\citet{10.1145/3351095.3372873} frames such concerns as an ``accountability gap'' between system development and deployment, where such systems are generally not evaluated until after they have been deployed and have potentially ``already negatively impacted users.''
Many instances of related work have focused on gender and ethnic biases in generative systems, such as findings outlined in \citet{10.1177/00178969241274621}, where ``generated images included a disproportionately high proportion of white male medical students'' compared to the actual population.
Likewise, the study of DALL-E in \citet{10.1145/3649883} demonstrates pervasive gender and racial biases across depictions of occupations.
These issues are not limited to image generation either, as \citet{10.1038/s41562-023-01716-4} illustrates language-specific biases as well.
Hence, strong academic consensus exists concerning generative AI's potential ``to reproduce, exacerbate, and reinforce extant human social biases'' \citep{10.1145/3649883}.

To date, however, few studies have examined how these systems are biased in the kind of \emph{emotional} content they produce.
Worryingly, \citet{doi:10.1073/pnas.1320040111} shows how negative emotions spread in online spaces, such that increasing the supply and exposure to negative emotional content can impact the psychological well-being of those exposed \emph{and} downstream audiences.
This effect and others concerning emotion in the modern information environment are complex, covering a wide variety of fields.
Journalism has reckoned with the role of emotion in its field, as \citet{10.1111/soc4.12677} describes an erosion of traditional objectivity-focused journalism in the face of crisis reporting, and \citet{10.1080/21670811.2019.1697626} further advances this shift as part of a reaction to the new media environment, where audiences are generally more emotionally engaged.
Similarly, political science has long engaged with the role of emotion in politics, with multiple competing theories driving continued research \citep{10.3389/fpos.2022.1080884}.
Understanding the intersection of emotion and political science in the modern, online information environment has increasingly garnered attention as well, with new findings on the role of emotion in information quality and mobilization.
\citet{10.1177/1065912918786805.2019}, for example, finds online audiences are particularly mobilized to engage in protest when exposed to images of their friends and images that evoke enthusiasm and fear.

Further, studies of information quality in online spaces increasingly point to the role of emotion in mediating consumption of high-quality content.
We have strong evidence that emotion drives online engagement with political content \citep{10.1126/sciadv.ade9231}, and political elites benefit from highly hostile and emotionally engaging content \citep{10.1073/pnas.2024292118.2021}.
\citet{10.1186/s41235-020-00252-3} further demonstrates how emotionality increases susceptibility to uptake of low-quality content.

Given these biases toward and effects of negative emotions, our hypothesis that generative AI systems trained using the vast volumes of online, digital-trace data would, as \citet{10.1145/3649883} suggests, ``reproduce, exacerbate, and reinforce'' this anti-social bias.
Understanding---and by extension controlling---this bias is critical to advancing solutions to the AI accountability gap \citet{10.1145/3351095.3372873} describes, while also necessary for enhancing the quality and resiliency of our information space.


\section{Methods and Data}

\subsection{Datasets for Images, Emotion, and AI Prompts}

For model finetuning and initial evaluation, we leverage the EmoSet dataset in \citet{10.48550/arxiv.2307.07961}. 
Each of the 118,102 images in the multi-class dataset is annotated as one of 8 discreet emotions: Amusement, Awe, Contentment, Excitement, Anger, Disgust, Fear, or Sadness.
To evaluate our central hypothesis of whether generative AI systems are biased towards negative emotions, we also leverage the DiffusionDB dataset \citep{wang2022diffusiondb}, a text-to-image prompt dataset containing 14 million images generated by Stable Diffusion with prompts from real users. 
We evaluate the salience of emotions within both the prompts and images contained in the 2m-first-10k subset of this dataset, a collection of 10,000 prompt-image pairs.

\subsection{Emotion Identification in Images}
To identify emotions in AI-generated images, we evaluated three approaches: zero-shot vision-language models, fine-tuned image classifiers, and an auto-captioning pipeline in which images are first converted to text descriptions before applying text-based emotion classifiers. Zero-shot models (BLIP, CLIP, and ALBEF) established a baseline for how well general-purpose vision-language models could recognize emotional content without task-specific training. Auto-captioning approaches underperformed even the zero-shot models, suggesting a bottleneck in cross-modal transfer.We then fine-tuned three high-performing vision architectures—Google’s Vision Transformer (ViT), Microsoft’s Swin Transformer, and ConvNeXT—on the EmoSet dataset to determine which architecture best captured nuanced emotional cues. Model performance was evaluated using macro-average precision, recall, and F1, providing a standardized comparison of accuracy and generalizability across emotion categories.

As shown in Table \ref{tab:emo_img}, across all methods, fine-tuned models outperformed zero-shot approaches, with Google’s ViT achieving the highest overall F1. These results confirm that task-specific fine-tuning substantially improves emotion recognition in visual data, while zero-shot approaches remain less reliable for this domain. Based on these findings, we selected the fine-tuned ViT model as the primary method for all subsequent analyses; details and full comparison metrics are included in the Appendix.


\subsubsection{Classification Performance and Validation}

\begin{table*}[htb]
\centering
\begin{subtable}{.32\linewidth}\centering\scriptsize
{\begin{tabular}{l r r r}

\hline
Model & Precision & Recall & F1 \\ \hline
\hline
BLIP-base & \textbf{0.4039} & \textbf{0.3275} & \textbf{0.3150} \\
CLIP-ViT-B16 & 0.3809 & 0.2417 & 0.1695 \\
ALBEF-base & 0.3459 & 0.2371 & 0.2033 \\ \hline

\end{tabular}}
\caption{Zero-Shot}\label{tab:emo_img_zero}
\end{subtable}%
\hspace{0.5em}
\begin{subtable}{.32\linewidth}\centering\scriptsize
{\begin{tabular}{l r r r}

\hline
Model & Precision & Recall & F1 \\ \hline
\hline
Google ViT & \textbf{0.7380} & \textbf{0.7313} & \textbf{0.7343} \\
MS SWIN & 0.7065 & 0.6932 & 0.6973 \\
ConvNeXT & 0.6865 & 0.6685 & 0.6740 \\ \hline

\end{tabular}}
\caption{Fine-Tuned}\label{tab:emo_img_ft}
\end{subtable}%
\hspace{0.5em}
\begin{subtable}{.32\linewidth}\centering\scriptsize
{\begin{tabular}{l r r r}

\hline
Model & Precision & Recall & F1 \\ \hline
\hline
GoEmotions  & \textbf{0.2552} & \textbf{0.2197} & \textbf{0.2154} \\
Paletz  & 0.1425 & 0.1808 & 0.1225 \\
SemEval  & 0.1525 & 0.2081 & 0.1406 \\ \hline

\end{tabular}}
\caption{Auto-Captioning}\label{tab:emo_img_cap}
\end{subtable}%
\caption{Image-based Emotion-Recognition Macro-Average Performance Across Learning Strategies (bolded results are highest). Results show fine-tuned computer vision models outperform zero-shot and caption-based methods by a substantial margin when predicting emotion classes in EmoSet \citep{10.48550/arxiv.2307.07961}. Auto-captioning methods use BLIP2 to generate captions and Demux-MEmo \citep{10.1109/icassp49357.2023.10095597} for text-based emotion extraction.}\label{tab:emo_img}
\end{table*}

To assess the validity of our highest-performing emotion-recognition model for visual media, we follow a method outlined in \citet{de2024using}.
We sample $n=200$ image-label pairs from our automatically labeled images and assess whether we agree (A), agree with doubt (D), or disagree (N) with the emotion label.
From these measures, we compute the acceptance rate as $(A + D) / (A + D + N)$.
For comparison, we also measure this acceptance rate against the original EmoSet dataset.
For EmoSet, we see an acceptance rate = $0.96$, and for Google's ViT, we see $0.86$, lower but still quite high.

\subsection{Emotion Identification in Text}

This research requires us to extract emotion distributions from the prompts in the DiffusionDB dataset, which we later compare with emotion distributions from the corresponding images.

We leverage three models from the \textbf{Demux-MEmo} models to classify text into discrete emotions due to its state-of-the-art ability to handle complex, multi-label emotion recognition tasks. These models were specifically designed to leverage the relationships between emotions, making it an ideal choice for classifying the emotional content present in text. We specifically test models pretrained on the Semeval-2018, Paletz, and GoEmotions benchmarks.

One of the core reasons for selecting Demux-MEmo is its explicit training in a multi-label setting, where more than one emotion may be present in a single piece of text. Unlike traditional models that treat each label independently, Demux-MEmo integrates the correlations between emotions, understanding that certain emotions often co-occur. For instance, emotions like fear and sadness are frequently linked, and recognizing such connections is critical for accurate emotion classification. This capability is particularly valuable in our work, where texts often convey complex emotional states, rather than single, isolated emotions.

Demux-MEmo also includes a regularization mechanism that leverages both global and local emotion correlations. By understanding these relationships, the model reduces the risk of making contradictory predictions. For example, if a text is classified as expressing both "joy" and "sadness," the model will use its learned correlations to resolve these conflicting predictions more intelligently than models without this capability. This level of sophistication is essential for our work, where subtle emotional nuances are common.


\subsection{Measuring Biased Emotionality Across Modalities}

This work's core objective is to assess potential biases in the emotions encoded into AI-generated images versus those present in the underlying prompts that create them; that is, we ask whether a systematic bias in emotion exists between prompts and the images they produce, such that an emotion present in the text is over- or under-represented in the resulting image.
To make this assessment, we leverage the substantial literature that has put forth methods and metrics for assessing potential biases---e.g., \citet{10.5555/3157382.3157469}, \citet{10.5555/3327144.3327272}, and \citet{9378025}.
\citet{10.1145/3278721.3278729} introduces a narrow definition of bias, ``a model contains unintended bias if it performs better for [elements] containing some particular [labels] than for [elements] containing others,'' which we alter for our emotion-propagation task: \emph{An \underline{image-generation} model contains unintended \underline{emotional} bias if it performs better for \underline{prompts} containing some particular \underline{emotion} than for \underline{prompts} containing others.}
We can relate ``performance'' here to \emph{recall} for a specific emotion, or the degree to which an image-generation model's output reproduces the dominant emotion present in its source, text-based prompt.
That is, emotional bias toward emotion $E$  manifests as significantly higher recall---the degree to which the emotion in the prompt propagates to or is recalled in the image.
As described in \citet{LangerEtAl:2024}, however, focusing only on this ``hit rate'' ignores potential response biases that emerge from overly ``responding'' with that particular emotion, thereby increasing false positives for that emotion; signal detection theory \citep{Stanislaw:1999aa} would interpret such results as a bias toward that emotion in that the generative model exhibits a lower decision threshold for that emotion.
Under the expectation that---since humans are more responsive to negative stimuli---the training data for image-generation models likely over-represents particular negative emotions, then we would expect a negative emotional bias to manifest as high having both high recall and a high false-positive rate for a given negative emotion.

To measure this possibility, we extract the emotions present in prompts provided in DiffusionDB \citep{wang2022diffusiondb}.
These emotions establish ``ground-truth'' labels for images produced from these prompts.
For each prompt, we then generate the corresponding image via a generative AI model (e.g., Stable Diffusion, GPT, etc.), extract the emotions evoked by this resulting image, and compare these image-based emotions to the text-based ground truth from that image's paired prompt.
This process produces a set of paired labels from which we can calculate standard classification metrics for each emotion.
To assess unintended emotional bias, we calculate the recall/hit rate/true positive rate (TPR) and false positive rate (FPR) for each emotion. 
TPR can be understood as, ``Among prompts that actually contain emotion \emph{e}, how often does the generated image also contain \emph{emotion}?'' 
FPR can similary be understood as measuring, ``Among prompts that do not contain emotion \emph{e}, how often does the image still contain emotion \emph{e}?''
We therefore define bias toward a particular emotion as an emotion for which both the TPR the FPR will be higher than for other emotions.

To assess variability in these per-emotion metrics, we use bootstrapping with 1,000 resamples of the text-/image-label pairs. 
For each resample, we construct per-emotion distributions for TPR and FPR metrics and report mean TPR/FPR values and 95\% confidence intervals.

\subsubsection{Addressing Cross-Modality Differences in Emotion Labels}

A complication in comparing emotion labels across modalities is that the datasets and models used for text- and image-based emotion recognition are not aligned.
That is, for EmoSet, the creators use the Mikels eight-category model~\citep{10.3758/bf03192732} to label emotion, but in the state-of-the-art text-based methods, emotion labels vary widely.
While we are consistent in using the Demux-MEmo model \cite{10.1109/icassp49357.2023.10095597} for predicting emotions in text, configuration for this model supports three separate datasets---SemEval 2018 Task 1, GoEmotion, and Paletz---each of which produces a different set of emotion labels. 
E.g., GoEmotion annotations cover 27 emotions, including``joy'' and ``surprise'', which are not present in the EmoSet labels; conversely, EmoSet includes ``awe'', which is not present in the GoEmotion  dataset.
Likewise, the SemEval dataset covers 11 emotions and a neutral class, with only ``fear'', ``sadness', ``anger'' and ``disgust'' explicitly in common with EmoSet.\footnote{One could argue that EmoSet's ``contentment'' might be equivalent to SemEval's ``joy'' class, but this alignment is not explicit in either  texts.}
The Paletz configuration for Demux-MEmo also has 23 emotions but still excludes ``awe'' and ``contentment'' from EmoSet.

While this misalignment exposes a clear path for future work, for this study, we follow \citet{demszky-etal-2020-goemotions} and map GoEmotions labels to those from the Ekman grouping \citep{ekman1992argument}.
This approach---while limited in that positive emotions are collapsed down to a single ``joy'' label---has the advantage of a consistent mapping in both modalities.
Table \ref{tab:ekman_map} specifies this mapping between modalities.
Then, for each of these mapped emotions, we calculate TPR/FNR and compare.

\begin{table}[htp]
\caption{Cross-Modality Ekman Mappings}
\begin{center}
\begin{tabular}{  l | p{0.67in} p{1.5in}}

& \textbf{EmoSet} & \textbf{GoEmotions} \\ \hline
 
 \textbf{Anger} & anger & anger, annoyance, disapproval \\\hline
 \textbf{Disgust} & disgust & disgust \\\hline
 \textbf{Fear} & fear & fear, nervousness \\\hline
 \textbf{Joy} & amusement, contentment, excitement & admiration, amusement, approval, caring, desire, excitement, gratitude, joy, love, optimism, pride, relief \\\hline
 \textbf{Sadness} & sadness & disappointment, embarrassment, grief, remorse, sadness \\\hline
 \textbf{Surprise} & awe & curiosity, realization, surprise, confusion \\ \hline

\end{tabular}
\end{center}
\label{tab:ekman_map}
\end{table}%



\section{Results}

\subsection{Estimating Bias in Image-based Emotion Recognition}

Before examining potential cross-modality biases, we first examine potential biases in our bespoke image-based emotion recognition model.
To this end, we leverage an external collection of images with emotion tags, namely the Emotion6 dataset~\citep{Peng_2015_CVPR}, and assess our emotion-recognition model's performance on identifying the human-provided emotion annotations therein.
The Emotion6 images, sourced from the Flickr image-sharing platform and hand-annotated using Amazon's Mechanical Turk platform, provide an out-of-sample assessment for our EmoSet-trained model.

\begin{figure}[htbp]
\begin{center}
\includegraphics[width=0.33\textwidth]{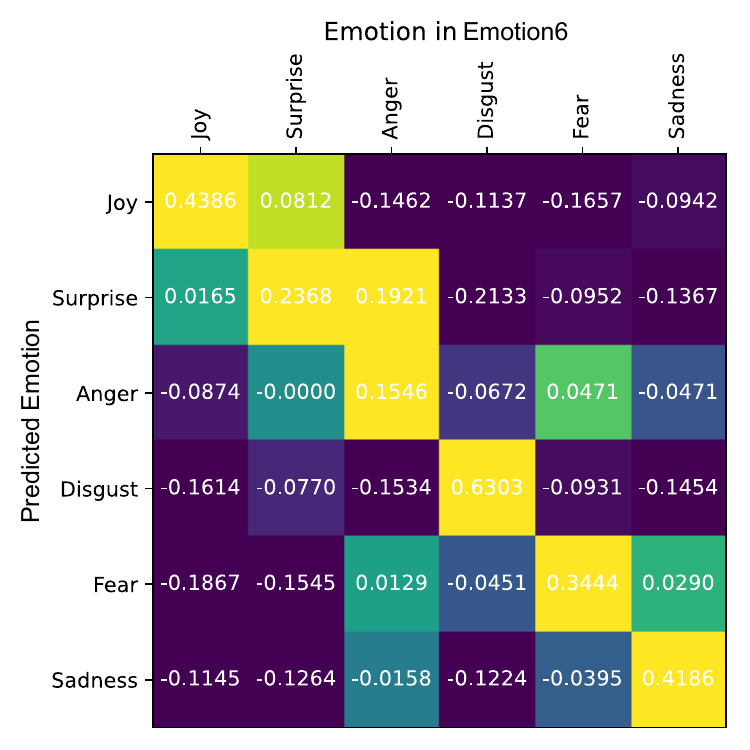}
\caption{Spearman $\rho$ Correlations in Ekman Emotions Between Emotion6 and Predicted Labels.}
\label{fig:emotion6_cross}
\end{center}
\end{figure}

For this dataset, Figure \ref{fig:emotion6_cross} shows the pairwise correlations between the Emotion6 emotions and our predicted emotions.
The on-diagonal correlations are the strongest between the Emotion6 labels and predicted emotions, though with the caveat that we see a higher than expected correlation between anger in Emotion6 and surprise as a predicted emotion.

\begin{figure*}[htbp]
    \centering
    \begin{subfigure}{.45\linewidth}\centering\scriptsize
    \includegraphics[width=0.75\linewidth]{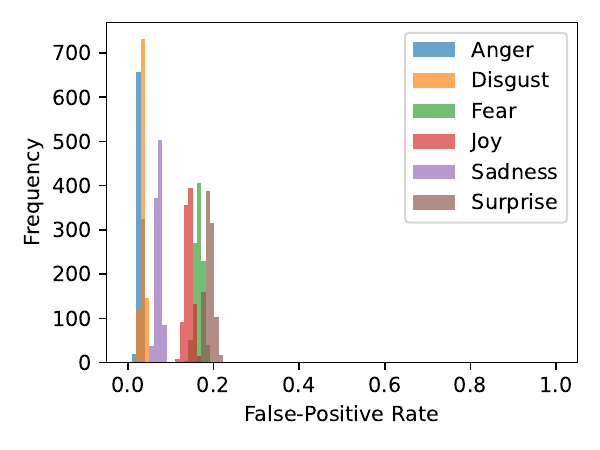}
    \caption{False-Positive Rate}
    \label{fig:image.emotion.fp_rate.emotion6}
    \end{subfigure}%
    \hspace{2em}
    \begin{subfigure}{.45\linewidth}\centering\scriptsize
    \includegraphics[width=0.75\linewidth]{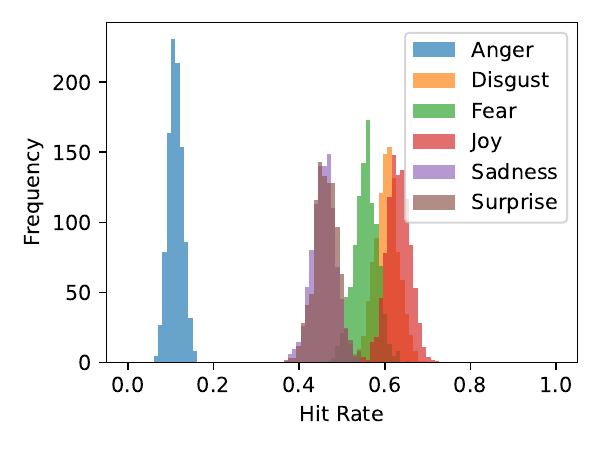}
    \caption{True-Positive Rate}
    \label{fig:image.emotion.hit_rate.emotion6}
    \end{subfigure}%
    \caption{Bootstrapped Performance Metrics for Classifying Emotions in the Emotion6 Image Dataset.}
    \label{fig:recognizing.emotion6}
\end{figure*}

To ensure potential emotional biases in the image-generation models are not attributable to biases in our recognition model, we first measure the per-emotion false- and true-positive rates against Emotion6, as shown in Figure \ref{fig:recognizing.emotion6}.
True-positive rates (Figure \ref{fig:image.emotion.hit_rate.emotion6}) are generally clustered, with means between 0.4 and 0.7, with joy, disgust, and fear having the top three highest true-positive rates; anger is a clear outlier here in that our model performs marked worse for anger classification in this dataset.
With the exception of anger, these results are generally in line with the convolutional neural network approach described in \citet{Peng_2015_CVPR}, which reports the true-positive rates for all emotions as between 0.55 and 0.7.
False-positive rates (Figure \ref{fig:image.emotion.fp_rate.emotion6}) see similar clustering---means below 0.2---with surprise fear and joy showing the highest false-positive rates.

In addition to these tests, we manually validate a subsample of image-emotion pairs extracted via our model and find high agreement between the human and learned model in the dominant emotion present in the image.
We also generate a variety of random images containing Gaussian noise, similar to \citet{Nguyen_2015_CVPR}, and classify these noisy images using our emotion-recognition model.
Results show that our model over-predicts joy in these images of random noise.

Taken together, these results suggest relatively minimal default bias toward recognizing any one emotion in an image for our image-based emotion-recognition model.

\subsection{Cross-Correlations Between Modalities}

\begin{figure*}[htbp]
\begin{center}
\includegraphics[width=0.95\textwidth]{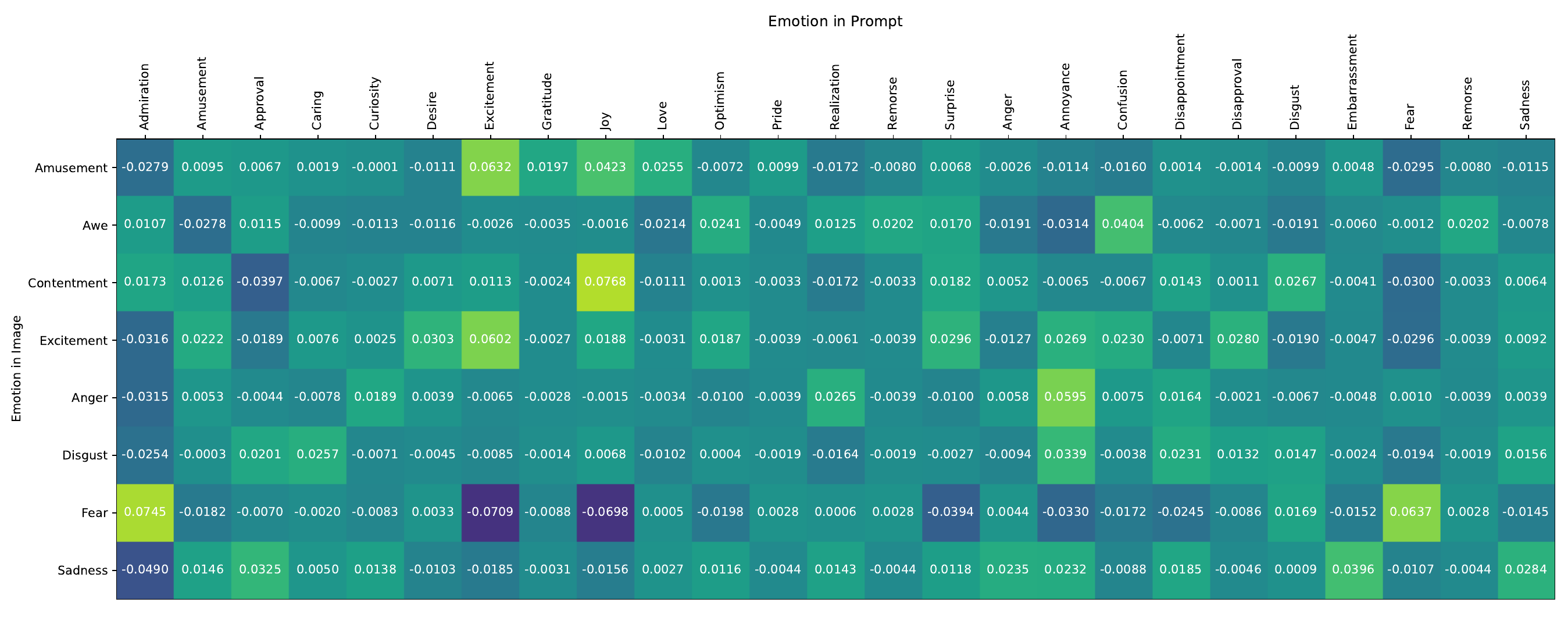}
\caption{Cross-Modality Spearman $\rho$ Correlations Between All Emotions in Prompts and Generated Images. }
\label{fig:demux_emoset_full_cross}
\end{center}
\end{figure*}

\begin{figure}[htbp]
\begin{center}
\includegraphics[width=0.33\textwidth]{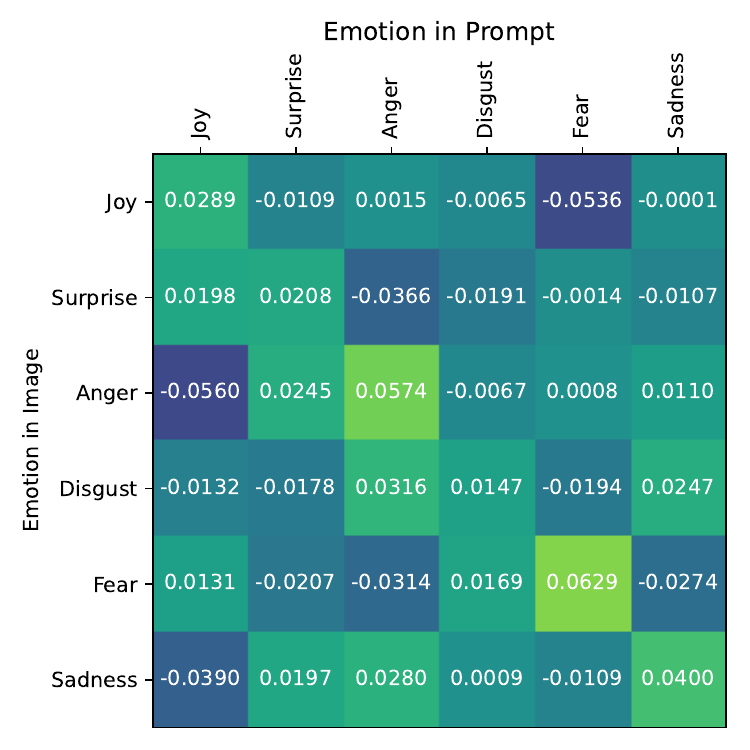}
\caption{Cross-Modality Spearman $\rho$ Correlations Ekman Emotions in Prompts and Generated Images. }
\label{fig:demux_emoset_cross}
\end{center}
\end{figure}

First looking across the full set of emotions extracted via text and images from images/EmoSet and text/GoEmotions (see Figure \ref{fig:demux_emoset_full_cross}), we see that, generally, emotions between the two modalities are, at best, only weakly correlated.
Upon aggregating to the six emotions in the Ekman model (Figure \ref{fig:demux_emoset_cross}). correlations among pairs of emotions increase: E.g., cross-modality correlation for sadness increases substantially, going from $0.0133$ to $0.0419$, and anger nearly triples from $0.0121$ before aggregation to $0.0395$ after).
Despite this increase, overall correlations remain weak, with fear having the maximum of cross-modality correlation of $0.0609$ and average cross-modality correlation of only $0.0327$.
For context, the scale of correlations seen here---$[-0.1, +0.1]$---is unsurprising given that \citet{Paletz:2024aa} shows correlations even within the same modality are in the 0.3 range.

\subsection{Comparing Emotions in Prompts and Generated Images}

\begin{figure*}[htbp]
    \centering
    \begin{subfigure}{.45\linewidth}\centering\scriptsize
    \includegraphics[width=0.75\linewidth]{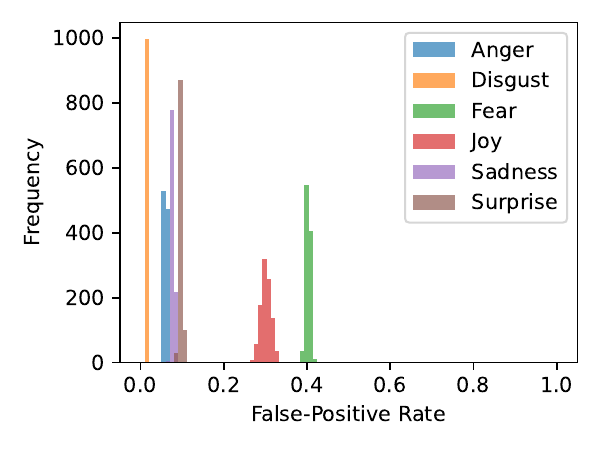}
    \caption{False-Positive Rate}
    \label{fig:xmodal.emotion.fp_rate}
    \end{subfigure}%
    \hspace{2em}
    \begin{subfigure}{.45\linewidth}\centering\scriptsize
    \includegraphics[width=0.75\linewidth]{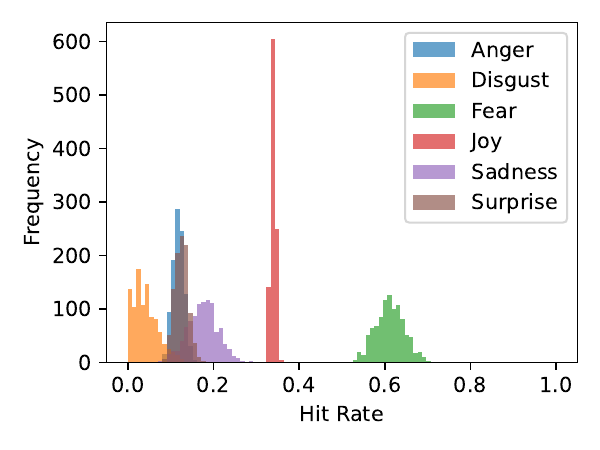}
    \caption{True-Positive Rate}
    \label{fig:xmodal.emotion.hit_rate}
    \end{subfigure}%
    \caption{Bootstrapped Performance Metrics for Consistency in Emotion Across Modalities. Both metrics show  these models exhibit a bias towards fear, as that emotion shows both the highest FPR and TPR, demonstrating a propensity toward evoking fear in the resulting image regardless of the underlying prompt.}
    \label{fig:xmodal.emotion}
\end{figure*}

Turning to the performance metrics for cross-modal preservation of emotions, results show that fear has both the highest FPR (Figure \ref{fig:xmodal.emotion.fp_rate}) and the highest TPR (Figure \ref{fig:xmodal.emotion.hit_rate}).
These results on TPR suggest that, if a textual prompt evokes fear, there is a better-than-even chance (approximately 60\%) that the resulting image will also evoke fear.
For all other emotions, there is a less than 50\% chance the resulting image will preserve the original dominant emotion.
FPR results also show that, on average, if an image evokes fear, a near-40\% chance exists that this emotion was not dominant in the underlying prompt, suggesting an overall bias toward producing images evoking fear despite the source text.

TPRs for the other emotions---particularly sadness, disgust, surprise, and anger---is low, with joy being the only other emotion that is preserved across modalities more than 1/3 of the time.
Notably, the FPRs for other emotions, with the exception of joy, are also low (less than 10\%), demonstrating that, if an image evokes sadness, disgust, surprise, or anger, it very likely came from a text-based prompt in which that emotion was dominant.
Taken together, these results are consistent with a generative model with a lower decision threshold for generating fear-adjacent imagery, potentially the result of a training dataset in which fearful imagery is over-represented, as we hypothesize.

\subsection{Assessment in Newer and Enterprise Models} 

The results above are specific to the Stable Diffusion model used to generate images in the DiffusionDB database, and while we find evidence of default-bias toward fear in those generated images, an open question remains about whether this bias is systemic or model-specific.
As the underlying mechanisms that may produce a training dataset of emotionally biased imagery are not model-specific, we hypothesize that this bias should exist in multiple models, including both more recent versions of Stable Diffusion and in closed-source models like OpenAI's GPT-Image-1.5 model.
To investigate this possibility, we sample 3,000 prompts from the 10,000 prompt-image-pair DiffusionDB sample used above and use OpenAI's Batch API to generate images from these prompts.
After completion, this batch image generation produced 1,642 1024x1024 images.
Unexpectedly, nearly half the prompts resulted in errors driven by OpenAI's content-moderation policies: Many prompts in the DiffusionDB sample included specific people's names, and these appear to have triggered moderation rules around likeness and intellectual property protections.
 
\begin{figure*}[htbp]
    \centering
    \begin{subfigure}{.45\linewidth}\centering\scriptsize
    \includegraphics[width=0.75\linewidth]{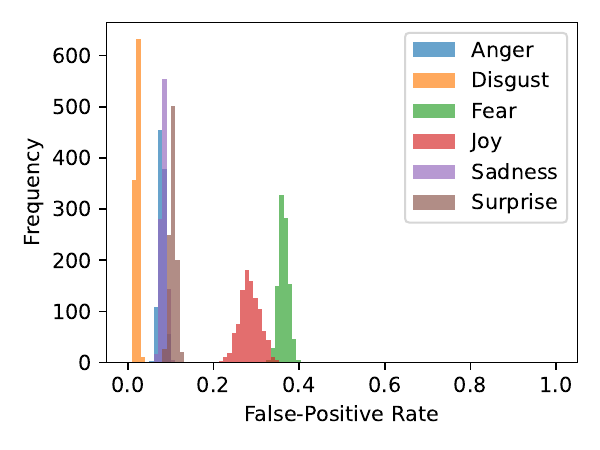}
    \caption{False-Positive Rate}
    \label{fig:xmodal.emotion.fp_rate.gpt}
    \end{subfigure}%
    \hspace{2em}
    \begin{subfigure}{.45\linewidth}\centering\scriptsize
    \includegraphics[width=0.75\linewidth]{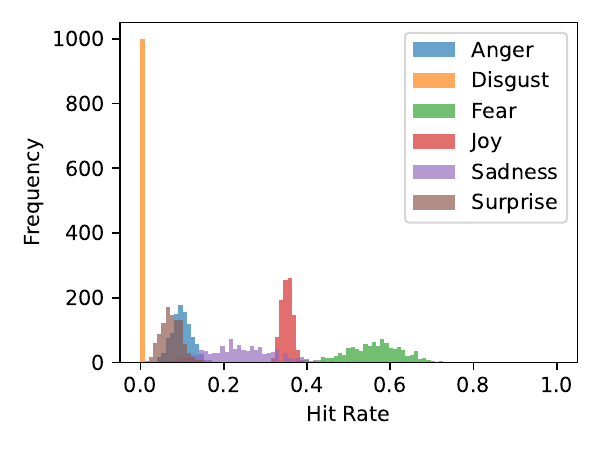}
    \caption{True-Positive Rate}
    \label{fig:xmodal.emotion.hit_rate.gpt}
    \end{subfigure}%
    \caption{Bootstrapped Performance Metrics for OpenAI's GPT-Image-1.5 Model. Results are consistent with the Stable Diffusion-based DiffusionDB results, as fear similarly has the highest true- and false-positive rates.}
    \label{fig:xmodal.emotion.gpt}
\end{figure*}

For GPT-Image-1.5, our Google ViT-based emotion-recognition model applied to these images produces the false- and true-positive rate distributions shown in Figure \ref{fig:xmodal.emotion.gpt}.
As illustrated in comparing Figures \ref{fig:xmodal.emotion} and \ref{fig:xmodal.emotion.gpt}, these bootstrapped distributions are highly similar in order: Fear has the highest rates for both metrics in both models, disgust has the lowest, and joy is consistently the second highest.
These results indicate a similar bias toward fear even in this closed model.

\begin{figure*}[htbp]
    \centering
    \begin{subfigure}{.45\linewidth}\centering\scriptsize
    \includegraphics[width=0.75\linewidth]{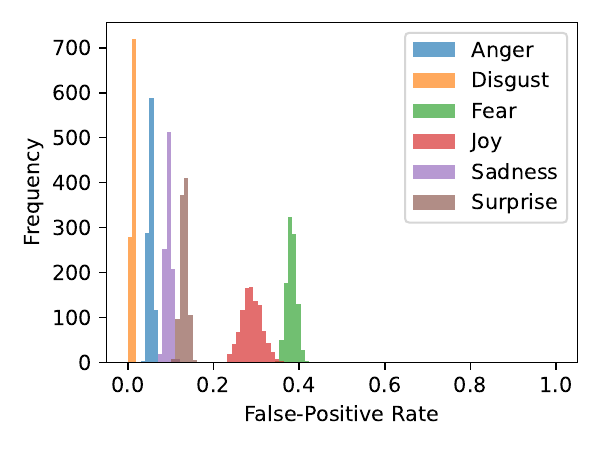}
    \caption{False-Positive Rate}
    \label{fig:xmodal.emotion.fp_rate.sdxl}
    \end{subfigure}%
    \hspace{2em}
    \begin{subfigure}{.45\linewidth}\centering\scriptsize
    \includegraphics[width=0.75\linewidth]{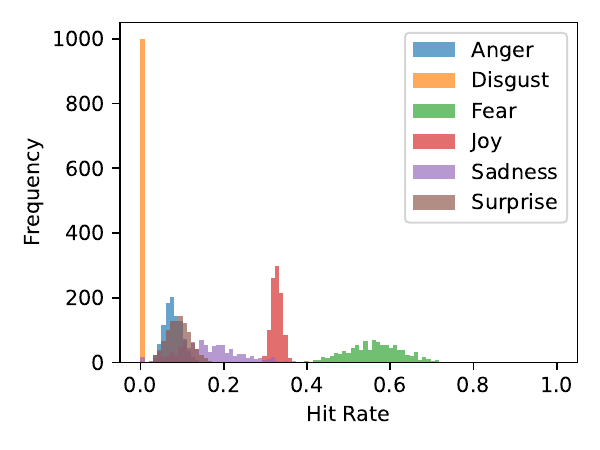}
    \caption{True-Positive Rate}
    \label{fig:xmodal.emotion.hit_rate.sdxl}
    \end{subfigure}%
    \caption{Bootstrapped Performance Metrics for SDXL. Results mirror DiffusionDB and GPT-Image-1.5.}
    \label{fig:xmodal.emotion.sdxl}
\end{figure*}

Several years after the release of Stable Diffusion \citep{rombach2022high}, the authors followed up with Stable Diffusion XL (SDXL), a larger and more sophisticated model for high-resolution image generation \citep{podell2024sdxl}.
Using the same 1,642 prompts from DiffusionDB that successfully passed GPT-Image-1.5 generation, we similarly generate images using this newer model and compare its false- and true-positive rates to that of these original Stable Diffusion and GPT models.
Figure \ref{fig:xmodal.emotion.sdxl} shows these distributions, and we note their marked similarity to GPT-Image-1.5 and the DiffusionDB results: All three show both the highest true-positive and false-positive rates for fear, followed by joy.
Unlike the other two models, SDXL shows more separate in false-positive rates, with surprise, sadness, anger, and disgust all reasonably separated---whereas both Stable Diffusion and GPT-Image-1.5 exhibits substantial overlap in false-positive rate for surprise, anger, and sadness.

\section{Discussion and Limitations}

\subsection{Unvirtuous Cycles and Emotional Bias}

While work from affective computing, psychology, communications, AI, and journalism have all studied various aspects of emotions, emotional inducement, and emotionality in imagery, the modern information ecosystem presents challenges at the intersection of these spaces.
In isolation, findings from HCI show images tend to increase engagement with content in online spaces, findings from psychology show emotional elicitation varies by modality~\cite{10.1177/1754073917749016}, and still other findings show journalistic sources' preference for violent visuals are both influence by and can impact real-world loss of life~\cite{10.1080/10584609.2014.880976}.
Likewise, studies of generative AI systems show clear non-conscious racial and gender biases\cite{10.3390/sci6010003,10.1016/s2589-7500(23)00246-7} .
Given the feedback loop between humans and the AI systems governing our tools and online spaces, one should anticipate large, generative AI systems that are trained on extensive collections of digital trace data will reflect similar biases.

Our results find evidence for such emotional biases in the automated production and generation of visual media.
These findings persist in images generated by both Stable Diffusion and by DALL-E, suggesting a result endemic to generative models rather than an isolated incident; naturally, a larger sample of models is needed to examine this possibility further, but these two methods are broadly available to the general public.
This availability, while valuable for democratizing content creation, increases the potential harms negative emotional bias may wrought. 
As \citet{doi:10.1073/pnas.1320040111} demonstrates, negative emotion is contagious in online spaces; this contagious is especially problematic in light of increasing trends of anxiety disorder in the US, especially among youths \citep{10.1016/j.jpsychires.2020.08.014}, with such harms especially impacting young women~\citep{Geiger:2019aa}.
These unintended consequences are not limited to emotional well-being and mental health either.
\citet{10.1177/1065912918786805.2019} also demonstrates the implications of negative imagery for political mobilization and instability, where exposure to negative---especially fear-inducing---visuals increased likelihood to protest.

\subsection{Breaking Unvirtuous Cycles}

We hope a main outcome of this work is to stimulate additional work in addressing emotional bias in generative AI systems and raising emotional bias to an important dimension of studies in bias.
Crucially, however, a first step in breaking this unvirutous cycle is raising \emph{awareness} of such emotional bias, as this awareness can empower users to acknowledge and account for its existence.
We need not rely solely on individual awareness though, as we can also adapt interfaces from  the fairness literature~\cite{patro2022fair} into new tools that empower users and designers to create more robust, intentional, and pro-social generative-AI systems.

\subsection{Limitations in Emotion Labeling: Mapping Across Diverse Emotion Sets}

EmoSet employs the Mikel model with eight categories, i.e., amusement, awe, contentment, excitement, anger, disgust, fear, sadness, where the former four are positive emotions and the latter four are negative ones. The Paletz model, however, contains 22 distinct emotions; GoEmotions contains 27 distinct emotions; and SemEval contains 11 emotions. Additionally, there are inconsistencies across different datasets and model configurations in terms of which emotions are even contained. GoEmotions, Paletz, and SemEval all lack awe and contentment, and SemEval furthermore lacks amusement and excitement.

In an ideal case, if discreet emotions corresponded 1 to 1 across modalities, we could compare directly and confirm if individual emotions were produced disproportionately in images relative to text. Since mappings aren't exact, we instead collapse to binaries (positive and negative emotions) and those discreet emotions which are shared to accurately compare miss-rates. We also utilize $\rho$ correlations as they are evaluated regardless of emotion synonymity and are left to further analysis by the reader.

\subsection{Limitations in Emotion Labeling: Multi-Class Versus Multi-Label}

This investigation has revealed a key limitation in the current datasets available, particularly in the  mismatch between the computational task's structure and the underlying psychological theory.
Namely, EmoSet's annotation process constrains every image to evoke a single emotion.
In this process, images first automatically receive an emotion label based on keywords associated with Mikels' eight-category model~\citep{10.3758/bf03192732}, retrieved during a web-scraping process. 
Human annotators then assess these labels for validity (rather than independently generating these labels manually). 
To finalize an image’s emotion label, at least 7 out of 10 annotators must agree. 
Although \citet{10.48550/arxiv.2307.07961} recognizes the inherent ambiguity of emotion recognition, this process still assigns only a single emotion to each image, creating a multi-label classification problem. 
Additionally, because annotators merely confirm emotions, they may assign an image to a certain emotion even if that emotion does not dominate the image’s overall emotional impact. For example, if an emotion evokes mostly contentment and some joy, but is automatically assigned the emotion of joy, annotators will confirm that the emotion evokes joy even while disagreeing that joy is the image's primary emotion.

While the EmoSet annotation process ensures consistency and scalability while paralleling existing emotion-recognition tasks from natural language process~\citep{chatterjee-etal-2019-semeval}, it oversimplifies the complexities of emotional evocation. 
Specifically, as outlined in \citet{Paletz:2024aa}, emotions are states that can be experienced and evoked simultaneously, and emotion annotation  should reflect this multi-label structure inherent to the task.
This difference in multi-class versus multi-label structure also has important implications for the modeling of emotion and its impact on social media engagement.

\begin{figure}
    \centering
    \begin{subfigure}{.45\linewidth}\centering\scriptsize
    \includegraphics[width=0.75\linewidth]{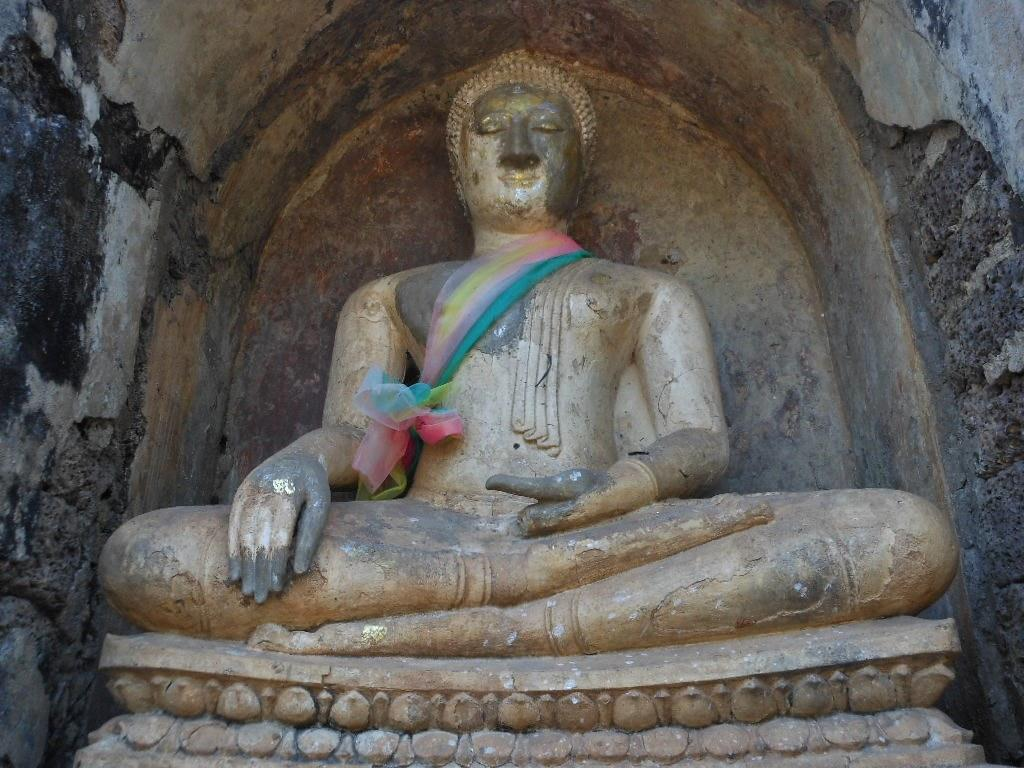}
    \caption{Image Awe\_07047, containing a smiling Buddha figure. Annotated as "Awe''.}
    \label{fig:awe-buddha-image}
    \end{subfigure}%
    \hspace{2em}
    \begin{subfigure}{.45\linewidth}\centering\scriptsize
    \includegraphics[width=0.4\linewidth]{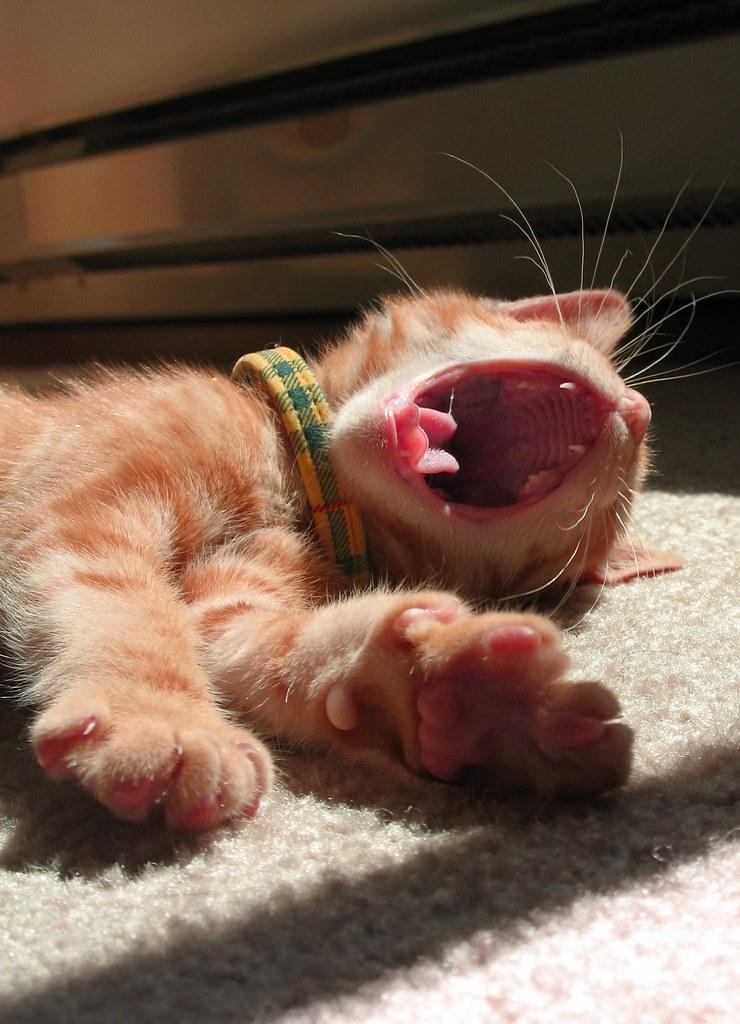}
    \caption{Image Anger\_07047, containing a stretching cat. Annotated as "Anger''.}
    \label{fig:angry-cat-image}
    \end{subfigure}%
    \caption{EmoSet Images and Single Emotion Labels.}
\end{figure}

We can also point to instances in the EmoSet dataset where such deviations are present, such as in Figure \ref{fig:awe-buddha-image}, which shows a statue of the Buddha.
The associated EmoSet label for this image is one of ``awe'', though depictions of the Buddha are often associated with themes of joy, contentment, awe, and other positive emotions. 
Similarly, Figure \ref{fig:angry-cat-image} shows a yawning cat, labeled ``anger'' in EmoSet, but this image may also evoke contentment, amusement, or cuteness/kama muta, the last of which a particularly important dimension in online spaces~\citep{10.3389/fpsyg.2023.1068373}. 
This issue is not new, as images have long been understood to evoke similar yet distinct emotions between various viewers, both in the computer vision space \citep{Peng_2015_CVPR} and in psychology \citep{10.1177/1754073917749016}---where different modalities have shown varying efficacy in emotional inducement.

This mismatch is not specific to EmoSet---as the foundation in \citet{10.3758/bf03192732} explicitly seeks to identify \emph{the most salient emotion} in a given image---but structuring the emotion recognition task as a multi-class one is problematic for two key reasons:

\begin{enumerate}
\item Images are known to evoke multiple similar and distinct emotions, and two annotators may disagree on which emotion is the most salient. Forcing one emotion then impacts the validity of the label.
\item Emotions are not easily distinguishable, as we see significant overlap between Amusement, Awe, Contentment, and Joy. Similarly, other emotion classifications beyond those used by EmoSet contain both positive and negative emotions with much overlap. Training models to identify singular emotions within images without considering overlap between alternate emotions raises similar validity concerns.
\end{enumerate}

\subsection{Limitations in Emotion Labeling: Lacking a ``Neutral'' Class}

Image classification models fine-tuned on EmoSet (i.e this paper's ViT, ConvNeXT, and SWIN models) treat multi-class image emotion classification as a zero-sum task and confidently project that a given image contains a significant amount of at least one emotion. In reality, many images may be neutral or may strongly evoke multiple separate emotions. Further complicating the matter is that EmoSet has no "neutral" classification for training purposes and similarly doesn't quantify how evocative a given emotion is. This creates skewed comparisons with the text emotion classification models, which identify multiple distinct emotions without treating the task as zero-sum. While the text models can identify relatively low amounts of multiple emotions within a single prompt, the image models are forced to identify high levels of only a single emotion, even if there are other emotions within the same image and/or the identified emotion is relatively minimal within the image because the image is primarily neutral.

\subsection{Future Work}



To address the limitations identified in sections 5.2 - 5.4, we call for the construction of a new, comprehensive image-emotion dataset. Each image should be annotated with a numerical rating for a set of distinct emotions; i.e any given image should have a rating 0-10 for amusement, awe, contentment, etc. Treating emotions in images in this multi-label manner and furthermore quantifying the emotions should allow models to more comprehensively identify emotional salience and classify images with unclear salience.

\section{Conclusions}

Though this paper assesses and selects multiple methods for recognizing emotions in images and text, this work is in service of the larger question concerning the potential for an unvirtuous cycle emerging in the feedback loop between content created for/or online spaces and the content produced by generative AI systems.
Our analysis finds evidence that at least two generative AI models, one of which is at or near current state of the art, are likely to over-represent negative emotion in the images that they generate compared to the prompts from which these images were generated.
This bias towards negative emotion has substantial implications for the health of the online information environment and the emotional and psychological well-being of the users of these generative AI systems.
While these results may only be specific to Stable Diffusion and ChatGPT-4o/DALL-E, the popularity and ease of general access to these models may exacerbate this emotional biases.
Through this work, we hope to shed light on this potential unvirtuous cycle and anti-social feedback loop so that future work can establish new measures to correct for it and educate and empower users of these technologies.

\bibliographystyle{ACM-Reference-Format}
\bibliography{references}

\end{document}